**Assessing the role of the spatial scale in the analysis of lagoon biodiversity. A case-study on the macrobenthic fauna of the Po River Delta.**

Giovanna Jona Lasinio[1]
*Dipartimento di Scienze Statistiche, Sapienza Università di Roma, Rome, Italy*
Alessio Pollice
*Dipartimento di Scienze Economiche e Metodi Matematici, Università degli Studi di Bari Aldo Moro, Bari, Italy*
Éric Marcon
*AgroParisTech-ENGREF, Korou, France*
Elisa Anna Fano
*Dipartimento di Scienze della Vita e Biotecnologie, Università degli Studi di Ferrara, Ferrara, Italy*

**Abstract**
The analysis of benthic assemblages is a valuable tool to describe the ecological status of transitional water ecosystems, but species are extremely sensitive and respond to both microhabitat and seasonal differences. The identification of changes in the composition of the macrobenthic community in specific microhabitats can then be used as an "early warning" for environmental changes which may affect the economic and ecological importance of lagoons, through their provision of Ecosystem Services. From a conservational point of view, the appropriate definition of the spatial aggregation level of microhabitats or local communities is of crucial importance. The main objective of this work is to assess the role of the spatial scale in the analysis of lagoon biodiversity. First, we analyze the variation in the sample coverage for alternative aggregations of the monitoring stations in three lagoons of the Po River Delta. Then, we analyze the variation of a class of entropy indices by mixed effects models, properly accounting for the fixed effects of biotic and abiotic factors and random effects ruled by nested sources of variability corresponding to alternative definitions of local communities. Finally, we address biodiversity partitioning by a generalized diversity measure, namely the Tsallis entropy, and for alternative definitions of the local communities. The main results obtained by the proposed statistical protocol are presented, discussed and framed in the ecological context.

**1. Introduction**
Transitional waters, such as coastal areas, are highly heterogeneous ecosystems in relation to the high variation of chemical, physical, morphological, hydrodynamic and/or functional factors (Basset et al. 2013). These systems are characterized by high instability (Sousa et al., 2009) which is associated to the instability of their fresh water sources (e.g. floods, droughts of rivers/streams and sediment transport) and to the sea tide. In recent years, phenological shifts have been observed in the vegetation of these ecosystems in strictly aquatic areas (Viaroli et al., 2001), with the disappearance of rooted macrophytes (e. g. *Zoostera* and *Ruppia*) replaced by phytoplankton and macroalgae ephemeral (e.g. *Ulva*) (Raffaelli et al. 1998), and in limiting areas (shore and shallow waters), with the marked decrease or disappearance of *Phragmites* that were previously abundant in the riparian zones of European coastal areas with relatively low salinity (Van der Putten, 1997; Fogli et al. 2002). These changes are associated to the increase of anthropogenic activities over the coastal areas such as fishing (fish, mussels and clams), sand abstraction, agricultural pollution by nutrients causing eutrophication (Viaroli et al. 2001), etc.. Thus, modification of the aquatic vegetation is expected to change the whole community and therefore the ecosystem functionality and provided ecosystem services (Eire and Ferguson, 2002; Smith 2003; Newton et al. 2014). The aforementioned aspects indicate the need to increase our knowledge on transitional areas' functioning in order to improve the development of biodiversity conservation plans. Legislation and actions have been adopted to stop further deterioration and restore these areas to a healthy state. The Water Framework Directive (WFD, 2000/60/EC) requires EU Member States to assess the ecological status of each water body in Europe and to ensure a sustainable management such that good ecological quality of all water bodies would be obtained by 2015. The analysis of benthic assemblages is a valuable tool to describe the ecological status of these transitional ecosystems, since macrobenthic fauna is known to be highly correlated with the sediment, which accumulates the multiple sources of organic enrichment and pollution (Pearson and Rosenberg, 1978). Macrobenthic species are extremely sensitive and respond to both microhabitat and seasonal differences, adding complexity to the variability of these ecosystems (Carvalho et al 2011). Consequently, the importance of habitat definition has often been highlighted in connection with lagoon ecological status assessment (Gamito et al. 2012).

---

1 Corresponding author, DSS, Sapienza University of Rome, P.le Aldo Moro 5,00185 Rome, Italy. Giovanna.jonalasinio@uniroma1.it

The identification of changes in the composition of the macrobenthic community in specific microhabitats can be used as an "early warning" for environmental changes which may affect the economic importance of lagoons, through their provision of ecosystem services (e.g. nutrient cycling, flood control, shoreline stabilization, water quality improvement, fisheries resources, habitat and food for migratory and resident animals and recreational areas for humans) (Basset et al., 2013; Pinna et al., 2013). Understanding assemblage responses to the environmental gradients at multiple spatial scales is an important issue in conservation biology (Bae et al., 2014). From a conservational point of view, several questions arise. What is the importance of the biodiversity of a single microhabitat with respect to the entire ecosystem? Which microhabitats contribute more to the entire ecosystem biodiversity? Is it possible to maintain biodiversity of the entire lagoon preserving only the most diverse microhabitats or should we care more about the conservation of ecosystem peculiarities? Should we consider a lagoon as a combination of microhabitats? All these questions address the same fundamental issue, i.e. the appropriate definition of the spatial aggregation level of microhabitats or local communities (monitoring station, microhabitat, area, lagoon). A major focus of interest in Ecology has to do with understanding changes in patterns of diversity for spatial scales ranging from the local community to the entire ecosystem. Defining the local community corresponds to setting the spatial scale for the interactions between organisms and their environment. The partition between local and regional scales is an artificial idealization for a more complex reality. For most ecosystems there is a nested hierarchy of multiple spatial scales characterized by different biological processes (Loreau, 2000; Whittaker, 1977). The pressures affecting biodiversity patterns are often scale specific, making multiscale assessment a crucial methodological priority. As species richness is not additive, it is difficult to translate from the scale of measurement to the scale(s) of interest.

A number of methods have been proposed to tackle this problem, but some of them are too model specific to allow general application (Azaele et al. 2015): Brose et al. (2003) use simulated landscapes to examine the sensitivity of the bias and accuracy of different species richness estimators to spatial autocorrelation and strength of environmental gradients (among other things). Büchi et al. (2009) use a population-based model to simulate competing species in spatially explicit landscapes, investigating the influence of the spatial structure in habitat and disturbance regimes. Communities are here characterized by species richness and life-history traits. On the side of more widely applicable methods, Suurkukka et al. (2012) use multiplicative partitioning of true diversities to identify the most important scale(s) of variation of benthic macroinvertebrate communities across several spatial hierarchical scales. Azaele et al. (2015) introduce the spatial pair correlation function (PCF) to describe the spatial structure of species' abundances. PCF describes the correlation in species' abundances between pairs of samples as a function of the distance between them. Rajala and Illian (2012) introduce a family of spatial biodiversity measures by flexibly defining the notion of the individuals' neighbourhood describing proximity of locations within the framework of graphs associated to a spatial point pattern.

The main objective of this work is to assess the role of the spatial scale in the analysis of lagoon biodiversity, the focus being on one ecosystem and a relatively small geographical scale. We first analyze the variation in the sample coverage for alternative aggregations of monitoring stations in three lagoons of the Po River Delta. Then, we analyze the variation of a class of entropy indices by mixed effects models properly accounting for the fixed effects of biotic and abiotic factors and random effects ruled by nested sources of variability corresponding to alternative definitions of local communities. Finally, we address biodiversity partitioning by a generalized diversity measure, namely the Tsallis entropy, and for alternative definitions of the local communities.

The workflow of the paper is displayed in Figure 1. In Section 2 some technical details are given on the sampling procedures, statistical methods and mathematical ideas applied in this work. In Section 3 results are described and finally a discussion is provided in section 4.

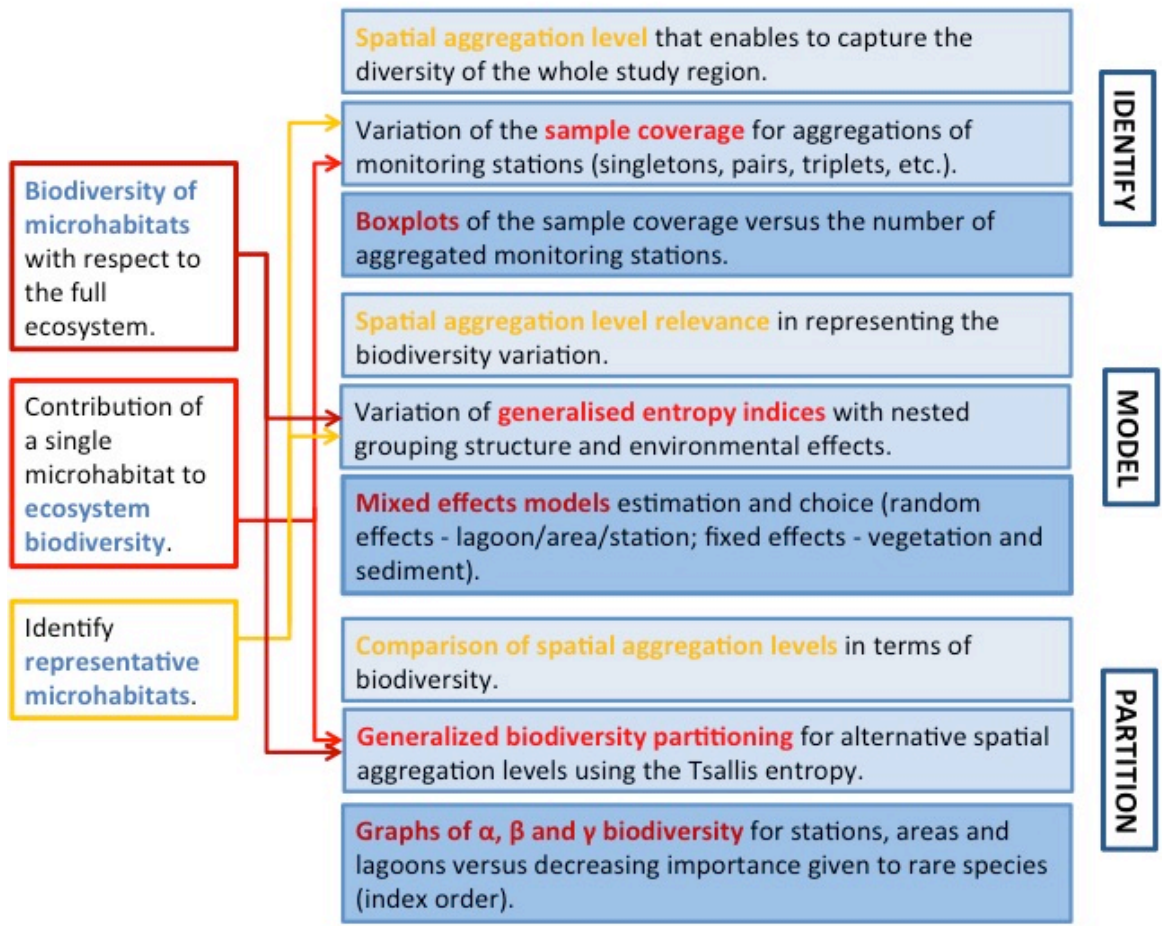

**Figure 1.** Conceptual scheme of the interactions between the ecological processes and the statistical protocol.

## 2. Materials and methods
### 2.1. Study Area
The study area is composed by three lagoons, all connected to the Po River Delta in Northern Italy and influenced by the Adriatic Sea. The Goro lagoon is a shallow-water embayment of the Po River Delta (44.78-44.83°N, 2.25-12.33°E), approximately triangular in shape with a surface area of 26 Km$^2$, an average depth of 1.5 m, and it is connected to the Adriatic Sea by two mouths about 0.9 Km wide each. The western area (Valle Giralda) is mainly influenced by freshwater inflow from the Po di Volano, while the northern part (Taglio della Falce) is influenced by both freshwater and brackish water. The central area (Valle di Goro) is mainly influenced by sea water. The eastern area (Valle di Gorino) is very shallow (maximum depth 1 m) and accounts for one half of the total surface area and one quarter of the water volume. The most southern part of the Goro lagoon is named Valle di Goro scanno, it is the closest to the open sea and bordered by sand dunes (further information in Viaroli et al. http://www.dsa.unipr.it/lagunet/infosheet/02-goro.pdf). The Comacchio lagoon, the largest lagoonal ecosystem in the Po River Delta, is located in the coastal area of the north-western Adriatic Sea, about 40 km south of Ferrara. The Valli di Comacchio is a semi-enclosed lagoonal complex of about 115 km$^2$, with an average depth of 1 m (0,5-1,5m), almost completely surrounded by earthen dikes, and separated from the sea by the highly anthropogenically impacted, 2.5-km wide Spina spit. This lagoonal system is connected with the Adriatic Sea by three marine channels, the Porto Canale, Logonovo, and Gobbino, but since the latter is impounded, exchange of the Valli occurs only through the Porto Canale and Logonovo channels. Both channels enter the lagoonal complex through Valle Fattibello, which also receives a small amount of water through the Navigable channel, and then flow into the northern basin of Valle Magnavacca. The complex of Comacchio includes six basins, five of them are considered in this study: Valle Magnavacca (about 62 km$^2$), Valle Campo (about 30 km$^2$), Valle Smarlacca (about 2 km$^2$), Valle Fattibello (about 8 km$^2$) and Valle Spavola (about 2 km2). Valle Fattibello and Valle Spavola, are separated from the others and have been considered as a lagoon by themselves named Fattibello. The bottoms of the Valli are typically muddy characterized by bare sediment (Valle Spavola and Valle Campo), or sparsely vegetated meadows of the seagrass *Ruppia cirrhosa* (Valle Magnavacca) or characterized by the presence of the green macroalgae *Ulva rigida* C. Ag. (Valle Fattibello). (Lagunet site: http://www.dsa.unipr.it/lagunet/infosheet/03-comacchio.pdf). Valle Smarlacca is located in the southeast corner of the Comacchio lagoon, close to the Reno River. It has a mean water depth of 0.8 m and the superficial sediment is mainly composed of organically enriched silts. This organic layer is 10-20 cm thick and overlies a deeper clay layer. The aquatic phanerogam *Ruppia cirrhosa* forms large patchy meadow, alternating between areas of dense canopy and areas devoid of plants. Salinity is relatively stable (22 to 24 psu) but can rise to 25-30 psu in summer due to evaporation. The area is surrounded by embankments and is completely separated from the other basins of the Comacchio lagoon. Valle Smarlacca receives freshwater and nutrient inputs from the adjacent Reno River through artificially-regulated sluices. The Valle Smarlacca area is also exploited for fish farming (Lagunet site: http://www.dsa.unipr.it/lagunet/infosheet/04-smarlacca.pdf).

### 2.2 Data Collection
Data on benthic macroinvertebrates were collected in three lagoons (Goro, Fattibello and Comacchio) within the Po River Delta ecosystem (Northern Adriatic Sea, Figure 2). The selected lagoons present from one to three dominant habitat types defined by a factorial classification of sediment granulometry (sand, mud) and vegetation cover/type (without vegetation, submerged macrophytes, emerged macrophytes and macroalgae) as in Basset et al. (2008b). According to this classification, every observed habitat was sampled with at least three replicates at each of the sites (monitoring stations) reported in Figure 2. Furthermore, to capture as much variability as possible, the sampling design accounted for salinity gradients and morphological specificities of each lagoon (for example confinement). The monitoring sites were sampled monthly in the period 1997-2000. At each site, within each individual lagoon, three replicate benthic samples were collected at the same place for the analysis of the macrofaunal community. The fauna retained on a 0.5 mm screen were identified to the lowest practical taxonomic level (usually species) and counted. The sediment at macroalgal beds and Manila clams bed stations was sampled in triplicate using a Van Veen grab (surface area 0.06 m$^2$) with a penetration depth of 12 cm. When present, the macroalgae were collected in triplicate, using a benthic hand grab net (mesh size 500 μm, mouth size 0.4 m), pulled for 1 m to cover an area of 0.4 m$^2$. Filtered sediment samples were stored in 1 L polyethylene bottles and fixed using buffered 8% formaldehyde. Macroalgal samples, mainly represented by pleustophytes *Ulva* sp. and *Gracilaria verrucosa* (Hudson) Papenfuss, and more rarely by *Cladophora* sp., were stored in plastic bags, within refrigerated containers and sorted when reaching the laboratory, in a few hours with the animals until alive. Bare sediment stations were sampled in triplicate, using plexiglass core liners (8 cm i.d.). *Phragmitetum* and *Ruppietum* stations were sampled in triplicate using steel squares (20x20 cm) mounting a 500 μm mesh net, randomly positioned on the sediment after the removal of reed stems and *Ruppia* and inserted in the sediment to a 20 cm depth. To collect the sample, roots underneath were cut using a shovel. The collected samples were filtered (500 μm) in situ and then stored 1 L polyethylene bottles, using 8% buffered formaldehyde.

A total of 47 taxa (see the **Appendix** for details) have been identified. Hereby, we present results including 3 replicates at each of 23 monitoring stations, divided in 10 areas belonging to 3 lagoons (nested grouping structure).

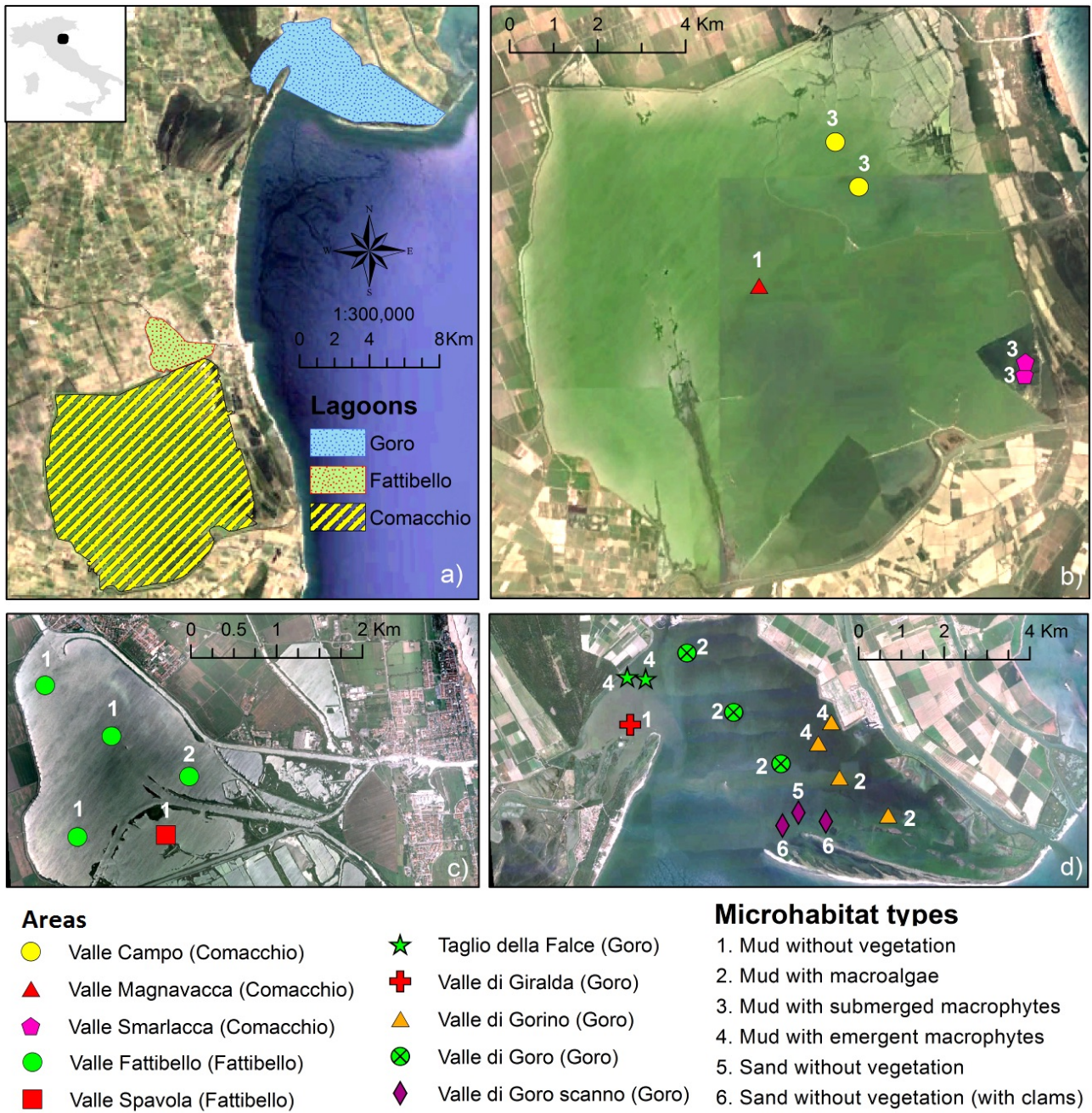

**Figure 2.** a) Location of the three lagoons, monitoring stations and microhabitat types in b) Comacchio, c) Fattibello and d) Goro (notice that the three areas Valle di Giralda, Valle di Goro and Valle di Goro scanno can be referred to as Valle Giralda, Sacca di Goro and Sacca di Goro scanno, as in Ludovisi et al., 2013).

**2.3 Statistical Protocol**
We address the main ecological questions as summarized in the introduction and sketched in Figure 1 by a specific statistical protocol structured in the three steps corresponding to general and specific objectives as described below.

**Identify -** A first grasp of the spatial aggregation level that captures the diversity present in the Po River Delta is obtained by the sample coverage of combinations (single, pairs, triplets, etc.) of monitoring stations. Given $OS_{obs}$ the number of observed species and $Q_1$ the number of species which occurred in just one of $m$ sampling units (uniques), as in Burnham and Overton (1979) we define the Jackknife 1 estimator of the number of observed species by:

$$OS_{est} = OS_{obs} + Q_1 \cdot \frac{m-1}{m}$$

The sample coverage, following Brose et al.'s (2003) definition, is the proportion of sampled species: $SC = OS_{obs}/OS_{est}$. Brose et al. (2003) show that the Jackknife 1 estimator becomes accurate if the sample coverage is above 74%.

In order to assess the number of stations whose data must be aggregated to reach a sufficient accuracy of richness estimation, stations are combined in all possible ways, up to the maximum number available in each lagoon and the sample coverage is obtained for each combination. Boxplots of the sample coverage are then plotted versus the number of stations considered in the combinations. Graphical representations of these numbers help to evaluate the "optimal" number of stations to consider at once. Generally speaking, we expect that the estimated number of species and the sample coverage tend to "stabilize" increasing the number of stations considered in the combinations. To be conservative, we fix a threshold of 80% for the sample coverage.

**Model -** Mixed models, also known as mixed effects models or multilevel models, find application when the data have some sort of grouped and/or hierarchical structure. In the analysis of biodiversity, grouped data are provided by repeated measurements within the same communities, while nested data structures arise due to hierarchically scaled local/spatial communities. If the experimental design implies taking multiple measures per subject (e.g. monitoring station, habitat, community), then sample data can be considered as made of groups of individual responses. This leads to a violation of the usual independence assumption implied by the linear model, as multiple or repeated measures from the same group cannot be regarded as independent from each other. In mixed models random effects are added to the systematic terms or fixed effects. Adding a random effect for each group of a grouped data structure allows to resolve the non-independence issue and to characterize variations due to group differences by a different random baseline for each group. Fitting mixed effects models implies the selection of relevant fixed and random effects. The latter is generally based on tools like the Akaike Information Criterion (AIC) and the Bayesian Information Criterion (BIC), (see for example Burnham, Anderson, 2002) both made of two terms measuring the fit and the complexity of the model, respectively. The maximized model likelihood is used as measure of fit and can be obtained by two alternative approaches: simple likelihood maximization (ML) and restricted expected likelihood maximization (REML). AIC's and BIC's based on ML are not comparable with those obtained by REML. A complementary approach to model selection is via hypothesis testing, with three options: t-statistics, F-statistics or likelihood ratio tests. Mixed model selection is generally based on a top-down procedure starting with the so called *beyond optimal model*, where the fixed component contains all explanatory variables and as many interactions as possible. The optimal random structure can be found comparing BIC and AIC values of the REML estimates of the beyond optimal model with alternative nested specifications of the random structure. Then the F-statistic or the t-statistic can be used to find the optimal fixed structure. To compare models with nested fixed effects, but with the same random structure, ML estimation must be used. The final estimates of the selected model are obtained using REML (see Zuur et al., 2009 pages 116-122 for details on the REML-ML differences and use in mixed models).

Multilevel mixed effects models accounting for fixed and nested sources of variability are used here to analyze the variation of generalized entropy indices corresponding to a set of selected levels of spatial aggregation:

$$h_{rmls} = \beta_1 x_{1,rmls} + \beta_2 x_{2,rmls} + \beta_3 x_{3,rmls} + b_s + b_{ls} + b_{mls} + \varepsilon_{rmls}$$

where $h_{rmls}$ is the value of the generalized entropy index, with $r = 1, \ldots, 12$ for months, $m=1,\ldots,23$ for monitoring stations, $l = 1, \ldots, 10$ for areas and $s = 1,2,3$ for lagoons. The previous model accounts for the fixed effect of the season (factor $x_1$), of the vegetation ($x_2$) and of mud and sand sediment types ($x_3$). Random effects of the lagoon ($b_s$), of the area ($b_{ls}$) and of the monitoring station ($b_{mls}$) reflect the nested grouping structure of the local communities. After model selection and fit, estimates are interpreted and allow finding, among other features, which random effects are more influential in explaining the variability of generalized entropy indices, then returning evidence on the relevance of the spatial aggregation, i.e. the habitat size.

**Partition** - We analyze several biodiversity measures in relation to the definition of the spatial aggregation level or local community. We partition gamma diversity measures into alpha and beta diversity and compare across spatial aggregation levels. The aim is here to understand the influence on biodiversity measures of the relevant spatial aggregations highlighted in the "model" step. Information theory and entropy measures have extensively been applied to ecological problems in areas as different as biodiversity assessment, evolution, species interactions and landscape analysis. Within the common focus of measuring ecosystem structural and functional complexity, Zaccarelli et al. (2013) deal with the temporal dynamics of normalized spectral entropy measures, Ludovisi and Scharler (2017) frame such measures in the context of ecological network analysis, Butturi-Gomes et al. (2017) highlight the statistical relevance of the Tsallis measure in comparison to several entropy function used as biodiversity indices.

Consider a community where $n$ individuals are sampled. Let $s = 1, \ldots, S$ denote the species that compose the community and $n_s$ be the number of sampled individuals of species $s$, with $\sum_{s=1}^{S} n_s = n$. The probability that an individual belongs to species $s$ is estimated by $p_s = n_s/n$. Given a discrete set of probabilities $p = (p_1, \ldots, p_S)$ and any real number $q$, the Tsallis entropy of order $q$ (Marcon and Hérault, 2015a, 2015b), is defined as

$$H_q(p) = \frac{1}{q-1}\left(1 - \sum_{s=1}^{S} p_s^q\right) \quad (1)$$

The number of species minus 1 is the Tsallis entropy of order $q = 0$, while Shannon's and Simpson's indices respectively correspond to $q \to 1$ ($q = 1$ in what follows) and $q = 2$, then the importance given to rare species decreases continuously with $q$. Corresponding *true diversity* measures $D_q(p)$, or *Hill numbers*, are obtained taking the deformed exponential transformation $e_q$ of the Tsallis entropy:

$$D_q(p) = e_q(H_q(p)) \quad (2)$$

For $H_q(p) < \frac{1}{q-1}$, the deformed exponential transformation of order $q$ is defined as

$$e_q(H_q(p)) = [1 + (1-q)H_q(p)]^{\frac{1}{1-q}} \quad (3)$$

with the standard exponential transformation obtained as a special case when $q = 1$.

Diversity measures are traditionally partitioned into gamma, alpha and beta diversity, respectively $^\gamma D_q(p)$, $^\alpha D_q(p)$ and $^\beta D_q(p)$, with $^\gamma D_q(p) = {^\alpha D_q(p)} \times {^\beta D_q(p)}$. Biodiversity partitioning means that the gamma biodiversity of all individuals in a given meta-community may be split into alpha and beta biodiversity that respectively reflect the diversity within and between local communities. The multiplicative partition of biodiversity measures is mathematically equivalent to the additive decomposition of the Tsallis entropy, as stated in Marcon et al. (2014) where the authors define beta entropy as the average generalized Kullback-Leibler divergence between local communities and their average distribution. They also propose estimation bias corrections that can be applied to the Tsallis entropy to obtain, after deformed exponential transformation, easy-to-interpret components of biodiversity. Output interpretation is helped by graphical representation of the diversity components as functions of the order $q$, together with 95% bootstrap confidence bands (Efron and Tibshirani, 1986). We do expect that if spatial aggregation is uninfluential, then curves corresponding to different spatial aggregation levels tend to overlap.

## 3. Results

A first simple procedure to find out the aggregation level that captures the diversity present in the Po River Delta consists in obtaining the fraction of all species found or "sample coverage" for single monitoring stations, pairs, triplets, etc.. We collect all such combinations up to the maximum number of stations that can be aggregated in each lagoon. We expect that increasing the number of aggregated stations most of the combinations return the same sample coverage. In Figure 3 we report boxplots of the sample coverages ($SC$), observed number of species ($OS_{obs}$) and number of uniques ($Q_1$) obtained for each combination of monitoring stations in the three lagoons. Analyzing Figure 3 we see that each lagoon shows its specific behavior with respect to sample coverage: Fattibello satisfies the 80% threshold with any number of stations, while Comacchio requires at least 2 stations to fully capture the diversity of the lagoon. Goro is the largest water body among those considered and has the largest number of uniques, thus it requires at least 5 stations to exceed the threshold.

In the following we consider generalized entropies as computed in equation (1) for $q$ = 0, 1, 2. Notice that the following identities are implied by equation (3):

$$H_0(p) = D_0(p) - 1$$
$$H_1(p) = \log D_1(p)$$
$$H_2(p) = 1 - \frac{1}{D_2(p)}$$

To assess which spatial scale (i.e. definition of the local communities: lagoons, areas or monitoring stations) is more relevant in representing the entropy variation, we estimate mixed effects models with fixed effects defined by season, vegetation and sediment indicators describing the surrounding environment. Random effects are ruled by the nested effects lagoon/area/monitoring station.

When the beyond optimal model is estimated with vegetation classified into 4 categories (none, macroalgae, emerged macrophytes, submerged macrophytes), the choice between alternative random structures highlights the relevance of the lagoon and station effects, while the area effect appears to be negligible (with AIC for $q$ = 0, 1, 2, with BIC for $q$ = 0, 1). AIC and BIC comparisons of models with the selected random structure and alternative specifications of the fixed effects leads to ambiguous results for $q$ = 0, 1, 2. The previous considerations (results are not reported here for the sake of space and readability, but are available from the authors upon request) pushed towards a more aggregate definition of the information on the vegetation coverage, considering the presence/absence of macroalgae. In Table 1 the "beyond optimal" approach introduced in section 2 for the choice of random effects is reported, with vegetation coded as macroalgae presence/absence. The most relevant feature highlighted is that the intermediate spatial aggregation level (area) is generally not relevant for the description of entropy variability, while the hierarchical combination of lagoon and station effects returns the best value (according to all chosen criteria, except for $q$ = 2 with the BIC).

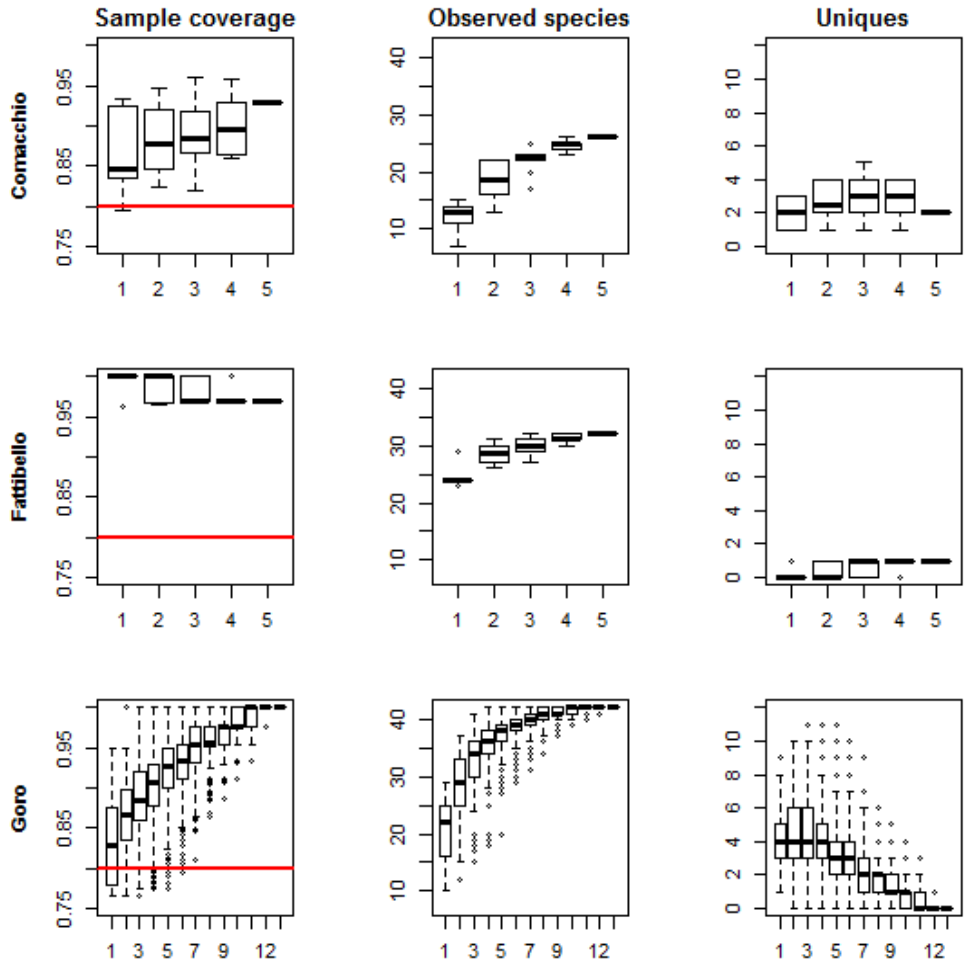

**Figure 3.** Assessing the number of stations required to capture the species in each lagoon boxplots of the sample coverages ($SC$), observed number of species ($OS_{obs}$) and number of uniques ($Q_1$) obtained for each combination of monitoring stations in the three lagoons

| Random effects | q=0 | | q=1 | | q=2 | |
|---|---|---|---|---|---|---|
| | BIC | AIC | BIC | AIC | BIC | AIC |
| lagoon/area/station | 1477.213 | 1430.775 | 394.6358 | 348.1978 | -86.4343 | -132.8723 |
| lagoon/area | 1498.456 | 1455.590 | 396.7004 | 353.8345 | -90.1588 | -133.0246 |
| lagoon/station | <span style="color:red">1471.641</span> | <span style="color:red">1428.775</span> | <span style="color:red">389.6561</span> | <span style="color:red">346.7902</span> | -90.8121 | <span style="color:red">-133.6780</span> |
| area/station | 1484.811 | 1441.945 | 393.9559 | 351.0900 | -88.2805 | -131.1463 |
| lagoon | 1494.840 | 1455.547 | 400.3243 | 361.0306 | -88.8064 | -128.1001 |
| area | 1508.481 | 1469.187 | 396.0933 | 356.7996 | <span style="color:red">-91.9428</span> | -131.2365 |
| station | 1493.181 | 1453.887 | 396.9530 | 357.6593 | -87.0491 | -126.3428 |

**Table 1.** Choice of the random effects based on both AIC and BIC criteria, for 3 different biodiversity indices: q=0 number of species, q=1 Shannon, q=2 Simpson. In red the chosen combination.

In Table 2 we report the choice of the fixed effects with the nested random effects lagoon/station chosen above. Again the choice is based on both BIC and AIC criteria. Macroalgae presence together with the type of sediment seem to play a crucial role no matter how we measure biodiversity. Season seems to have a marginal relevance only when rare species are given a relevant role (q=0) and according to only one criterion (AIC). In summary the model structure is the same for the three specifications of the biodiversity index order ($q = 0, 1, 2$) including the nested random effects of monitoring stations within lagoons and the fixed effects of the presence of macroalgae and sediment. AIC and BIC help in choosing the "best" model, but they tell us little about how well the models fit the data or whether there are any departures from model assumptions. Here we concentrate on some graphical analyses of model performances. In Figure 4 the boxplots of residuals by lagoons show that the models chosen according to BIC and AIC criteria comply with the standard assumption of Gaussian error for the three orders of the generalized entropy measure. Normal q-q plots are useful for assessing whether the error distribution is modelled correctly and to detect more general departures from model assumptions (Zuur et al., 2009). As shown in Figure 5, the Gaussian distributional assumption also holds for the random effects of monitoring stations. The comparison of observed and fitted values (Figure 6) provides some evidence of a better fit when $q = 0$, i.e. when larger importance is given to rare species. In this case a few outlying observations all come from the Fattibello lagoon characterized by the smallest number of uniques (few rare species). When $q = 1, 2$ the fit is not as good and the outliers are found in the three lagoons. We have to remind that our main interest is in testing the role of the spatial aggregation on the behavior of biodiversity measures. Hence the overall goodness of fit has a less central role than the significance of fixed and random effects. In Table 3 the estimated fixed effects together with their standard errors and p-values for the chosen models (Table 2) are reported. As both sediment and macroalgae are classified as binary, the model intercept represents the fixed effect of mud and no-macroalgae. The other coefficients are deviations from this intercept value and are significantly different from zero whenever the corresponding p-value is smaller than a fixed error threshold chosen to be 5% (0.05). Hence when $q=0$ all effects are statistically significant. In the presence of a sand sediment and no macroalgae the expected entropy level is obtained adding 6.7984 to the estimated intercept, while if the sediment is sand and macrolagae are present the value 12.6674 (5.864+6.7984) has to be added to the estimated intercept effect. When $q=1$ and 2 (and less importance is given to rare species) the contribution of the sediment becomes weaker (p-values increase), but it is still important.

| Fixed effects | q=0 | | q=1 | | q=2 | |
|---|---|---|---|---|---|---|
| | BIC | AIC | BIC | AIC | BIC | AIC |
| Season * Macroalgae + Sediment2 | 1486.813 | <span style="color:red">1443.543</span> | 366.0261 | 322.7565 | -132.0346 | -175.3042 |
| Season * Macroalgae | 1498.510 | 1458.846 | 366.5817 | 326.9179 | -131.5525 | -171.2163 |
| Season + Macroalgae + Sediment | 1478.230 | 1445.778 | 351.1506 | 318.6984 | -147.5589 | -180.0111 |
| Season + Macroalgae | 1489.923 | 1461.076 | 351.7047 | 322.8583 | -147.0784 | -175.9248 |
| Macroalgae + Sediment | <span style="color:red">1473.405</span> | 1451.770 | <span style="color:red">339.4166</span> | <span style="color:red">317.7817</span> | <span style="color:red">-159.9924</span> | <span style="color:red">-181.6272</span> |
| Season + Sediment | 1492.917 | 1464.071 | 358.8131 | 329.9666 | -138.9411 | -167.7875 |
| Season | 1490.706 | 1465.466 | 353.7882 | 328.5476 | -144.1383 | -169.3789 |
| Macroalgae | 1485.118 | 1467.089 | 339.9750 | 321.9460 | -159.5090 | -177.5381 |
| Sediment | 1488.114 | 1470.085 | 347.0858 | 329.0568 | -151.3701 | -169.3991 |
| Intercept | 1485.908 | 1471.484 | 342.0621 | 327.6389 | -156.5669 | -170.9901 |

**Table 2.** Choice of the fixed effects given the random effects lagoon/station. The choice is based on both BIC and AIC criteria, for 3 different biodiversity indices: q=0 number of species, q=1 Shannon, q=2 Simpson. In red the chosen combination.

| q=0 | Value | Std.Error | p-value |
|---|---|---|---|
| No Macroalgae and mud | 8.1057 | 3.1652 | 0.0110 |
| Macroalgae | 5.8640 | 1.0337 | 0.0000 |
| Sand | 6.7984 | 1.3401 | 0.0001 |
| **q=1** | | | |
| No Macroalgae and mud | 1.2586 | 0.2023 | 0.0000 |
| Macroalgae | 0.4825 | 0.1147 | 0.0005 |
| Sand | 0.3934 | 0.1488 | 0.0165 |
| **q=2** | | | |
| No Macroalgae and mud | 0.5800 | 0.0578 | 0.0000 |
| Macroalgae | 0.1762 | 0.0393 | 0.0003 |
| Sand | 0.1364 | 0.0510 | 0.0153 |

**Table 3.** Estimated fixed effects and their standard errors for the models chosen according to BIC and AIC criteria, for three levels of the generalized entropy measure. P-values refer to coefficient significance.

---

2 The * notation denotes interaction between two variables.

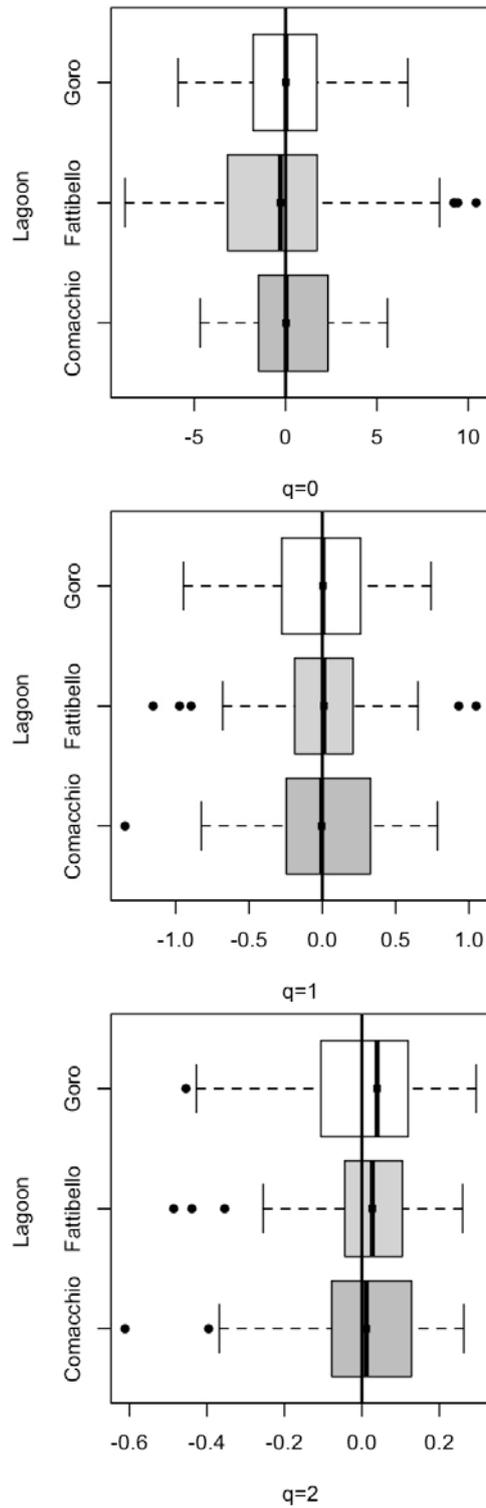

**Figure 4.** Boxplots of the residuals by lagoons for the models chosen according to BIC and AIC criteria and the three orders of the generalized entropy measure.

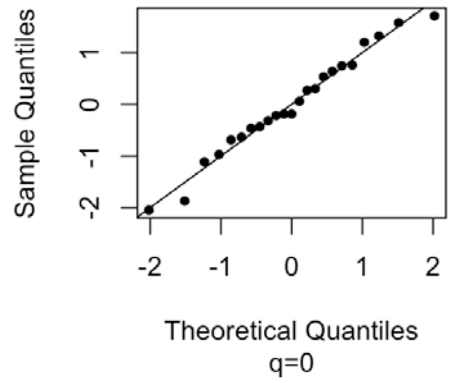

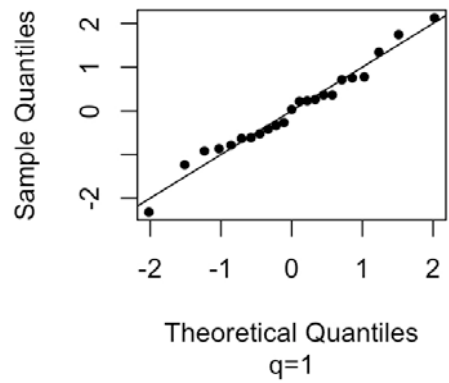

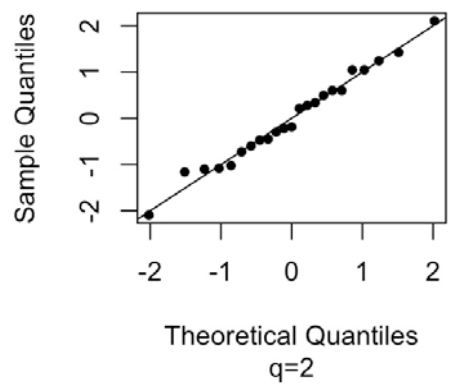

**Figure 5.** Normal q-q plots of the estimated random effects of monitoring stations for the models chosen according to BIC and AIC criteria and the three orders of the generalized entropy measure.

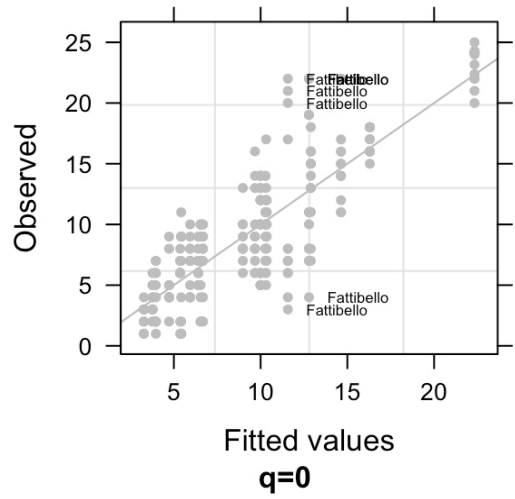

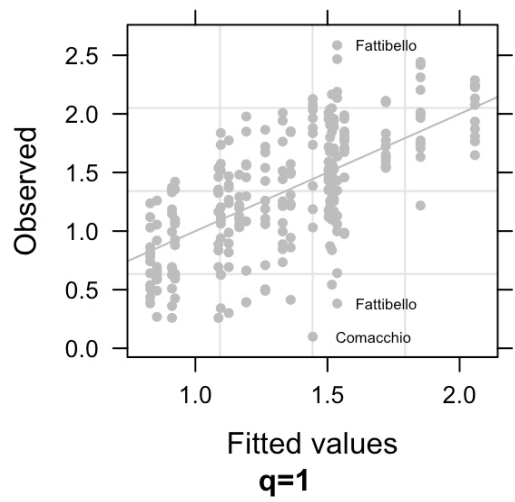

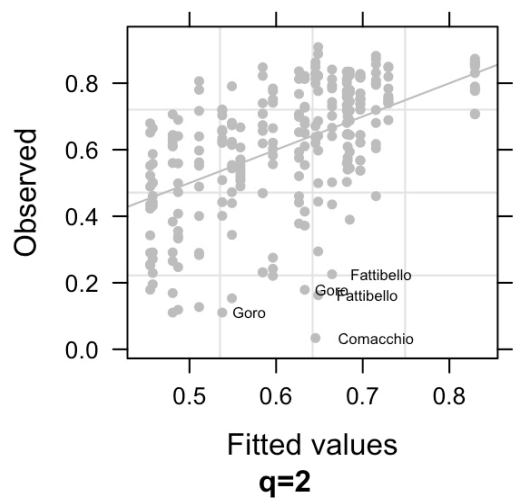

**Figure 6.** Observed vs. fitted values for the models chosen according to BIC and AIC criteria and the three orders of the generalized entropy measure.

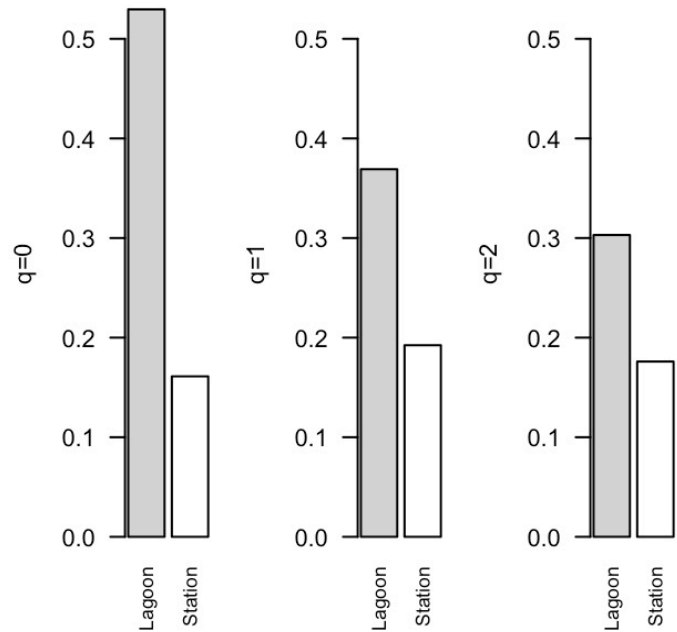

**Figure 7.** Rate of total standard deviation explained by the nested random effects (lagoon – blue, monitoring station – red) in the chosen models, for 3 levels of the order $q$ of the generalized entropy measure

Information on the relative relevance of the spatial scale of the local communities on the expected generalized entropy is obtained by the estimates of the random effects standard deviations. In Figure 7 the rates of the total standard deviation explained by each effect are depicted. The barplots show that the lagoon level is always the most relevant for these data, no matter the importance given to rare species. The role of the effect of monitoring stations becomes more important with respect to the one of lagoons for larger values of $q$, i.e. when rare species are less and less considered.

In Figure 8 we report the alpha, beta and gamma biodiversity of benthic macroinvertebrates for a range of values of the generalized biodiversity index of order $q$ and according to two different definitions of local communities: lagoons and monitoring stations. In particular we show biodiversity indices and their bootstrap 95% confidence bands distinguishing: (a) average alpha diversity of stations by lagoon, (b) gamma diversity of each lagoon, (c) beta diversity between lagoons, (d) beta diversity between stations by lagoon. In (a) we notice that Goro seems less diverse than the other two with an almost constant gap for $q > 0.5$. However these differences are not strongly supported given that the bootstrap confidence bands partially overlap for all values of $q$. Notice the remarkable higher biodiversity of Fattibello with respect to Comacchio for values of $q$ smaller than 0.5. Figure 8 (b) shows that the overall gamma diversity of Fattibello and Comacchio cannot be distinguished for any order of the index, while Goro has a significantly larger gamma diversity when $q$ is small. When $q$ reaches 0.5 the three water bodies become undistinguishable as the confidence bands overlap. The beta diversity among stations in the three lagoons (Figure 8 (d)) again suggests that they are not significantly different for $q>0.5$; only when q is close to 0 we observe a significant separation of the curves and their confidence bands: in particular larger difference among stations are obtained at Goro, then comes Comacchio and finally Fattibello. If eventually we look at the beta diversity among the three lagoons (Figure 8 (c)), a marked difference among the three water bodies is found when q is small with a considerable variability as shown by the width of the confidence bands.

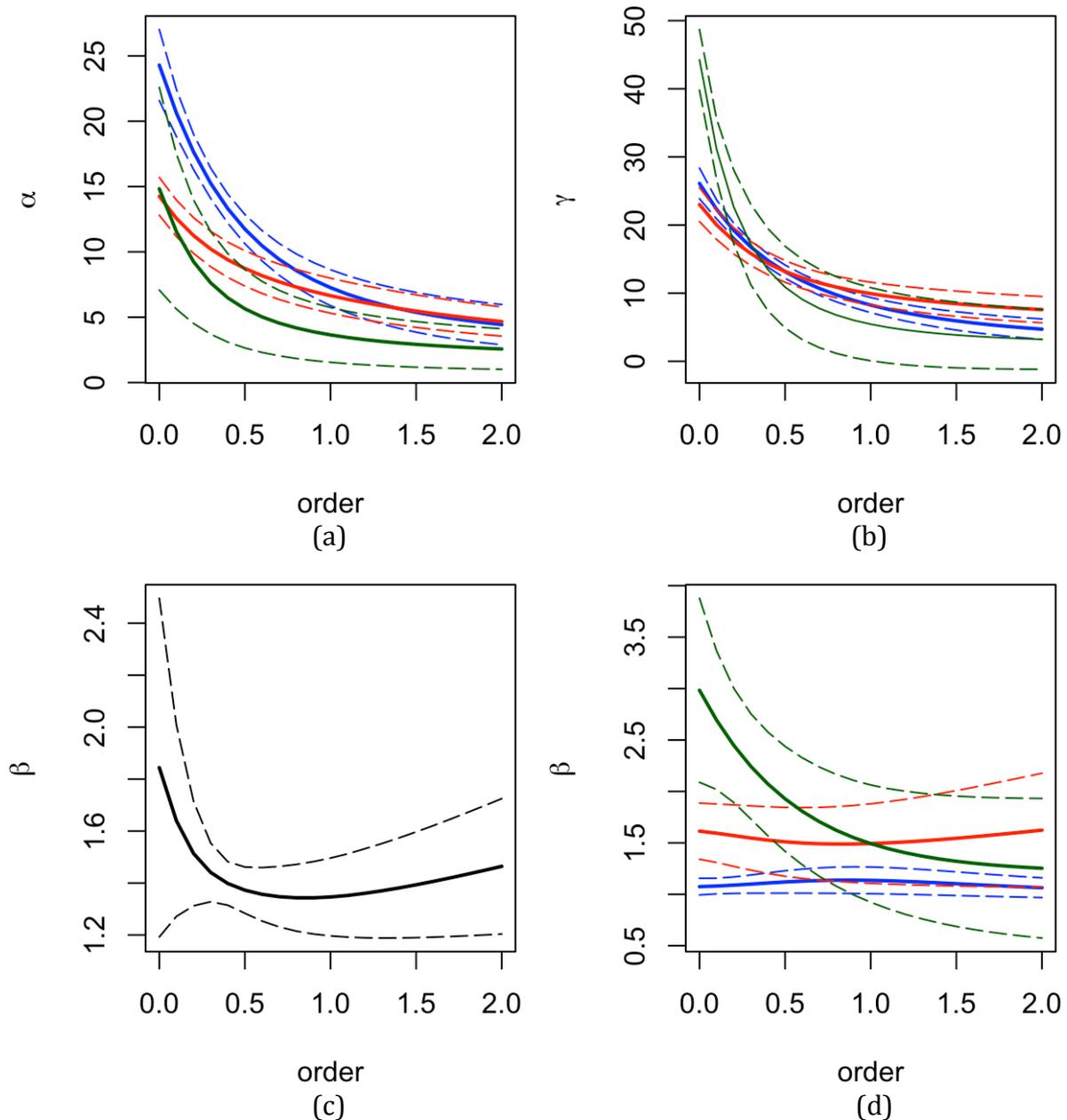

**Figure 8.** Biodiversity indices (solid) and their bootstrap 95% confidence bands (dashed) for the three lagoons (red – Comacchio, blue – Fattibello, green – Goro). (a) average alpha diversity of stations by lagoon, (b) gamma diversity of each lagoon, (c) beta diversity between lagoons, (d) beta diversity between stations by lagoon.

**4. Discussion**
The main findings obtained following the Identify step of the proposed protocol highlight that each lagoon shows a specific behavior with respect to the minimum number of stations required to fully capture its biodiversity. Namely, while Fattibello is fine with any number of monitoring stations, Comacchio requires at least 2 and Goro, the largest and most diversified water body, at least 5 stations. These findings are in line with the increasing degree of environmental heterogeneity acknowledged for Fattibello, Comacchio and Goro respectively.
Habitat heterogeneity has long been considered one of the determinants of biological diversity (Mac Arthur and Mac Arthur, 1961; Huston 1979) and is used as a predictor of species richness over a wide range of spatial scales, from individual to continental scales (e.g.

Gaston, 2000). Theoretically, the more complex a habitat the higher the number of species. Due in part to the technical difficulties of quantifying habitat heterogeneity, few studies have rigorously tested this hypothesis, but for example the relationship between habitat heterogeneity and macrozoobenthic diversity has been demonstrated in atoll lagoons of French Polynesia (Pante et al. 2006), and also in *Zoostera noltii* seagrass bed in the mediterranean Arcachon Bay (Blanchet et al. 2004).

The Model step of the proposed protocol allows to highlight several relevant features related to the influence of the spatial scale of microhabitats on the definition of the alpha diversity on the deformed log (i.e. entropy) scale. The hierarchical combination of lagoon and monitoring station effects is generally a relevant source of variability of the entropy measure, while the intermediate spatial aggregation level (area) has a poor performance in comparison. This highlights the possible role of the two habitat sizes (lagoon and station) as the main drivers of biodiversity variability over the three lagoons. Also the presence of macroalgae together with the type of sediment seem to play a crucial role, no matter how we measure biodiversity. When $q=1$ and 2 (and less importance is given to rare species) the contribution of the sediment becomes less important, while the season seems to have a marginal relevance only when rare species are given a relevant role ($q=0$).

These results confirm the main role of the vegetal community structure in influencing the macrozoobenthic component and determining differentiation in the community. It is well known that the composition of macrobenthic communities can vary on small spatial scales in transition systems (Ysebaert and Herman, 2002). This can be due to abiotic factors such as salinity, grain size of the substrate, the hydrology (Mannino and Montagna 1997; Barbone and Basset 2010), and biotic such as trophic relationships and the availability of food sources (Carvalho et al. 2011). Among the biotic factors, the vegetal community seems to deeply influence the structuring of the macrobenthic community. In particular, the presence of seagrass bed (Blanchet et al. 2004), macroalgae bed (Carvalho et all. 2011), emergent macrophytes (Yuhas et al. 2005) have the most relevant impact. The results of the estimate of mixed effects models also show that the geographical location doesn't contribute to this feature: analyzing geographically distant lagoons all characterized by the same seagrass bed structure leads to similar results most likely due to a strong homogeneity in the macrozoobenthic component. Macrobenthic communities accordingly respond to such mosaic-type lagoon environments structuring metacommunities with patches specific to available microhabitats. The degree of diversification reflects this structure, highlighting that all types of microhabitats play a relevant role in contributing to the diversification of the community as a whole (Davis et al., 2014).

The Partition step shows that, as expected, both diversity at the metacommunity level (gamma diversity of lagoons) and mean species diversity at the habitat level (diversity at the local community level or average alpha diversity of monitoring stations by lagoons) increase with the importance given to rare species in the biodiversity measure. Differentiation among microhabitats (beta diversity between stations by lagoons) shows the same feature only for the larger Goro Lagoon, also characterized by a higher number of species and by an overall lower average alpha biodiversity of monitoring stations. The Fattibello lagoon shows an opposite behavior: when the importance given to rare species is high, beta biodiversity is minimal and alpha biodiversity overrides the one of the other two lagoons. Indeed in the Fattibello lagoon rare species are not present (no singletons are found), this due to the high homogeneity of the lagoon microhabitat.

Going back to the questions we were asking in the introduction, this study is leading us to the following considerations:
- Is it possible to maintain biodiversity of the entire lagoon preserving only the most diverse microhabitats or should we care more about the conservation of ecosystem peculiarities? We found that the lagoon level is the one that explains the largest amount of variability regardless of the biodiversity measure considered. As local habitats are mainly caught at the station level, this suggests that we need to focus on the entire system and not on specific microhabitats or local features.
- What is the importance of the biodiversity of a single microhabitat with respect to the entire ecosystem? Analysis of the fixed model effects showed that while seasonal parameters are not significant, microhabitats indicators are. Effects estimates can then be used to evaluate the weight of microhabitats on the overall biodiversity measure. For example, in table 3 for q=0 we have an increase in biodiversity given by the theoretical combination of macroalgae and sand, the total contribution being 12.6624 of which 53.68% is given by sand.
- Which microhabitats contribute more to the entire ecosystem biodiversity? Should we consider a lagoon as a combination of microhabitats? Again, going back to table 3 we can say that the most diverse sites correspond to the simultaneous presence of macroalgae and sand, but the whole study points to the consideration of the entire lagoon as optimal reference for biodiversity assessment.

The proposed protocol facilitates understanding the influence of the spatial scale of the local communities on biodiversity. As a result of the three steps of the analysis, the differential behavior of the three lagoons of the Po River Delta ecosystem confirms the relevance of small-scale habitat structures highlighted in Hewitt et al. (2005) for coastal marine environments. The considerable functional specificity of the macrobenthic community with respect to food items presence can be seen as the main driver to small-scale analyses at microhabitat level (Barnes and Hendy, 2015). Results emphasize that also in the Po River Delta, as in other transitional water ecosystems (Basset et al. 2008a), biodiversity conservation cannot be managed locally but requires a large scale process of governance to be effective. This need to preserve higher levels of biological organization has long been recognized and is reflected in the emergence of a recent IUCN Red List of Ecosystems (Keith et al., 2013). In view of the conservation of transition systems, there is then the need for careful management of all types of microhabitat given their high value of specificity in term of both species and processes.

Possible alternative applications of the proposed methodology include understanding and quantifying patterns of biodiversity in landscape Ecology. Characteristics of landscape patterns as the spatial heterogeneity influence ecological processes and the resulting biodiversity with profound impacts on the functioning of ecological and socio-economic systems (Lausch et al., 2015). Methods to assess the geographic pattern of biodiversity are also called for in large cooperative studies of marine resources such as the COST action EMBOS (European Marine Biodiversity Observatory System), where the variation of the diversity of soft-bottom communities is investigated along with increasing latitude (Hummel et al., 2016).

**Acknowledgements**

Giovanna Jona Lasinio and Alessio Pollice were partially supported by the PRIN2015 project "Environmental processes and human activities: capturing their interactions via statistical methods (EPHASTAT)" funded by MIUR – Italian Ministry of University and Research. Anna Fano was partially supported by the RITMARE project, co-funded by MIUR – Italian Ministry of University and Research. She would like to thank many of her students who contributed to the data collection and laboratory analysis. Eric Marcon was funded by Agence Nationale de la Recherche (CEBA, ref. ANR-10-LABX-25-01. All the authors wish to thank Alberto Basset and Ilaria Rosati for the many helpful discussions.

# Appendix

| Taxa | Comacchio | | | Fattibello | | | Goro | | | | | | |
|---|---|---|---|---|---|---|---|---|---|---|---|---|---|
| | Valle Campo – mud, submerged macrophytes | Valle Magnavacca – mud without vegetation | Valle Smarlacca – mud, submerged macrophytes | Valle Fattibello – mud, macroalgae | Valle Fattibello – mud without vegetation | Valle Spavola – mud without vegetation | Valle di Goro – mud, macroalgae | Taglio della Falce – mud, emerged macrophytes | Valle di Gorino – mud, macroalgae | Valle di Gorino – mud, emerged macrophytes | Valle di Giralda – mud without vegetation | Valle di Goro scanno - sand without vegetation | Valle di Goro scanno - sand without vegetation with clams |
| Abra segmentum | 0 | 0 | 1 | 1 | 0 | 0 | 0 | 0 | 0 | 0 | 0 | 1 | 0 |
| Actiniari spp. | 1 | 0 | 1 | 1 | 1 | 1 | 1 | 1 | 1 | 0 | 0 | 0 | 1 |
| Alitta succinea | 1 | 1 | 1 | 1 | 1 | 1 | 1 | 1 | 1 | 1 | 1 | 1 | 1 |
| Ampelisca sp. | 0 | 0 | 0 | 0 | 0 | 0 | 0 | 0 | 0 | 1 | 0 | 1 | 1 |
| Anadara inaequivalvis | 0 | 0 | 0 | 1 | 1 | 1 | 1 | 0 | 1 | 1 | 0 | 1 | 0 |
| Arcuatula senhousia | 0 | 0 | 0 | 1 | 1 | 1 | 1 | 1 | 1 | 1 | 0 | 1 | 1 |
| Ascidiacei spp. | 0 | 0 | 1 | 1 | 1 | 0 | 1 | 0 | 1 | 1 | 0 | 0 | 1 |
| Balanidae spp. | 0 | 0 | 0 | 1 | 1 | 1 | 1 | 1 | 1 | 1 | 0 | 0 | 1 |
| Bittium reticulatum | 0 | 0 | 0 | 0 | 0 | 0 | 1 | 0 | 1 | 1 | 0 | 0 | 0 |
| Brachynotus sexdentatus | 0 | 0 | 0 | 1 | 1 | 1 | 1 | 0 | 1 | 0 | 1 | 1 | 1 |
| Capitella capitata | 1 | 1 | 1 | 1 | 1 | 1 | 1 | 1 | 1 | 1 | 1 | 1 | 1 |
| Carcinus aestuarii | 0 | 0 | 0 | 1 | 1 | 1 | 1 | 0 | 1 | 1 | 0 | 1 | 0 |
| Cerastoderma glaucum | 1 | 1 | 1 | 1 | 1 | 1 | 1 | 1 | 1 | 1 | 1 | 1 | 1 |
| Chironomus salinarius | 1 | 0 | 1 | 1 | 1 | 1 | 1 | 1 | 1 | 1 | 0 | 1 | 1 |
| Crassostrea sp. | 0 | 0 | 0 | 1 | 0 | 1 | 0 | 0 | 0 | 1 | 0 | 0 | 0 |
| Cyclope neritea | 1 | 0 | 1 | 1 | 1 | 0 | 1 | 0 | 0 | 0 | 0 | 1 | 1 |
| Ensis siliqua | 0 | 0 | 0 | 0 | 0 | 0 | 0 | 0 | 0 | 0 | 0 | 0 | 1 |
| Ficopomatus enigmaticus | 0 | 1 | 1 | 1 | 1 | 1 | 1 | 0 | 1 | 0 | 0 | 1 | 1 |
| Gammarus aequicauda | 1 | 1 | 1 | 1 | 1 | 1 | 1 | 1 | 1 | 1 | 1 | 1 | 1 |
| Haminoea hydatis | 1 | 0 | 0 | 0 | 0 | 0 | 1 | 0 | 1 | 1 | 0 | 1 | 1 |
| Heteromastus filiformis | 0 | 1 | 0 | 0 | 0 | 1 | 1 | 0 | 1 | 0 | 0 | 0 | 0 |
| Hydroides dianthus | 0 | 1 | 1 | 0 | 0 | 0 | 1 | 0 | 1 | 0 | 0 | 0 | 0 |
| Idotea balthica | 1 | 0 | 1 | 0 | 0 | 0 | 0 | 1 | 0 | 0 | 1 | 1 | 1 |

|  | Comacchio | | | Fattibello | | | Goro | | | | | | |
| --- | --- | --- | --- | --- | --- | --- | --- | --- | --- | --- | --- | --- | --- |
| **Taxa** | Valle Campo – mud, submerged macrophytes | Valle Magnavacca - mud without vegetation | Valle Smarlacca – mud, submerged macrophytes | Valle Fattibello – mud, macroalgae | Valle Fattibello - mud without vegetation | Valle Spavola - mud without vegetation | Valle di Goro – mud, macroalgae | Taglio della Falce – mud, emerged macrophytes | Valle di Gorino – mud, macroalgae | Valle di Gorino – mud, emerged macrophytes | Valle di Giralda - mud without vegetation | Valle di Goro scanno - sand without vegetation | Valle di Goro scanno - sand without vegetation with clams |
| Lekanesphaera hookeri | 0 | 0 | 0 | 0 | 0 | 0 | 0 | 1 | 0 | 1 | 0 | 0 | 0 |
| Lentidium mediterraneum | 0 | 0 | 0 | 0 | 0 | 0 | 1 | 0 | 0 | 1 | 0 | 3 | 0 |
| Littorina littorea | 1 | 0 | 0 | 0 | 0 | 0 | 0 | 0 | 0 | 0 | 0 | 0 | 0 |
| Lucifer typus | 0 | 0 | 1 | 0 | 0 | 0 | 0 | 0 | 0 | 0 | 0 | 0 | 0 |
| Microdeutopus gryllotalpa | 0 | 1 | 1 | 1 | 1 | 1 | 1 | 1 | 1 | 1 | 1 | 1 | 1 |
| Monocorophium insidiosum | 1 | 1 | 1 | 1 | 1 | 1 | 1 | 1 | 1 | 1 | 1 | 1 | 0 |
| Mytilaster minimus | 0 | 0 | 0 | 1 | 0 | 0 | 0 | 0 | 0 | 0 | 0 | 0 | 0 |
| Mytilus galloprovincialis | 0 | 0 | 0 | 1 | 1 | 1 | 1 | 0 | 1 | 0 | 0 | 1 | 0 |
| Nassarius reticulatus | 0 | 0 | 0 | 0 | 0 | 0 | 1 | 0 | 1 | 1 | 0 | 1 | 1 |
| Nephtys hombergii | 0 | 0 | 0 | 1 | 1 | 0 | 1 | 0 | 1 | 0 | 0 | 0 | 1 |
| Oligochaeta spp. | 1 | 1 | 1 | 1 | 1 | 1 | 1 | 1 | 1 | 1 | 0 | 1 | 1 |
| Ostrea edulis | 0 | 0 | 0 | 0 | 1 | 0 | 0 | 0 | 0 | 1 | 0 | 0 | 0 |
| Palaemon elegans | 1 | 0 | 0 | 1 | 0 | 1 | 1 | 0 | 0 | 0 | 0 | 0 | 1 |
| Phyllodoce lineata | 0 | 1 | 1 | 0 | 0 | 0 | 1 | 0 | 1 | 0 | 0 | 1 | 1 |
| Polydora ciliata | 0 | 1 | 1 | 1 | 1 | 1 | 1 | 1 | 1 | 1 | 0 | 1 | 1 |
| Prionospio multibranchiata | 0 | 1 | 1 | 1 | 1 | 1 | 1 | 0 | 1 | 1 | 0 | 1 | 1 |
| Ruditapes decussatus | 0 | 0 | 0 | 0 | 1 | 0 | 1 | 0 | 1 | 0 | 0 | 0 | 1 |
| Ruditapes philippinarum | 0 | 0 | 0 | 1 | 1 | 1 | 1 | 0 | 1 | 0 | 0 | 1 | 1 |
| Sphaeroma serratum | 0 | 0 | 1 | 0 | 0 | 0 | 0 | 1 | 0 | 0 | 0 | 0 | 1 |
| Spio decoratus | 0 | 0 | 0 | 0 | 0 | 0 | 0 | 0 | 1 | 1 | 0 | 0 | 1 |
| Streblospio shrubsolii | 0 | 1 | 1 | 1 | 1 | 1 | 1 | 1 | 1 | 1 | 1 | 1 | 1 |
| Tellina sp. | 0 | 0 | 0 | 0 | 0 | 0 | 1 | 0 | 1 | 1 | 0 | 0 | 1 |
| Turbellaria | 0 | 0 | 0 | 1 | 1 | 0 | 1 | 0 | 1 | 0 | 0 | 1 | 1 |
| Ventrosia ventrosa | 1 | 0 | 1 | 1 | 1 | 1 | 1 | 1 | 1 | 1 | 1 | 1 | 0 |

**Table A.** Presence (1) or absence (0) of species in Areas and type of microhabitat (Species nomenclature followed the World Register of MarineSpecies (WoRMS) 2016)

| **Main findings** | |
|---|---|
| **IDENTIFY** | Each lagoon has a specific number of stations **sufficient** to capture all the available species. |
| | Cautiously, Goro should be analyzed using 5 monitoring stations, 2 out of 5 are **enough** for the Comacchio lagoon, while Fattibello requires no more than 1. |
| **MODEL** | The hierarchical combination of lagoon and station effects returns the best value in terms of model fit, while the intermediate spatial aggregation level (area) is not relevant for the variability of the entropy measures. This highlights the possible role of **two habitat sizes** (lagoon and station). |
| | The presence of **macroalgae** together with the type of **sediment** seems to play a crucial role. When q=1 and 2 (and less importance is given to rare species) the contribution of the sediment becomes less important. The **season** seems to have a marginal relevance only when rare species are given a relevant role (q=0). |
| | The **lagoon** spatial aggregation level is the most relevant for assessing biodiversity in this system, no matter the importance given to rare species. The heterogeneity of **monitoring stations** becomes more evident when rare species are less and less considered. |
| **PARTITION** | **Diversity at the metacommunity level** (gamma diversity): Goro shows values higher than the other two for $q < 0.5$. |
| | **Mean species diversity at the habitat level** (alpha diversity): Fattibello shows values higher than Comacchio for $q < 0.5$. |
| | **Differentiation among microhabitats** (beta diversity): for $q < 0.5$ Goro and Fattibello respectively show higher and lower values with respect to Comacchio. |